\documentclass[submission,copyright,creativecommons]{eptcs}
\providecommand{\event}{SCSS 2021}

\usepackage{amsmath,amssymb,amsfonts}
\usepackage{algorithmic}
\usepackage[pdftex]{graphicx}
\usepackage{textcomp}
\usepackage{hyperref}
\usepackage{color,soul}
\usepackage{breakurl}   

\title{Statistical Model Checking of Common Attack Scenarios on Blockchain}
\author{
Ivan Fedotov \qquad\qquad Anton Khritankov
\institute{Moscow Institute of Physics and Technology\\
Moscow, Russia}
\email{\quad ivan.fedotov@phystech.edu \quad\qquad anton.khritankov@phystech.edu}
}
\def\titlerunning{SMC Blockchain Verification}
\def\authorrunning{Ivan Fedotov, Anton Khritankov}
\begin{document}
\maketitle

\begin{abstract}
Blockchain technology has developed significantly over the last decade. One of the reasons for this is its sustainability architecture, which does not allow modification of the history of committed transactions. That means that developers should consider blockchain vulnerabilities and eliminate them before the deployment of the system. In this paper, we demonstrate a statistical model checking approach for the verification of blockchain systems on three real-world attack scenarios. We build and verify models of DNS attack, double-spending with memory pool flooding, and consensus delay scenario. After that, we analyze experimental results and propose solutions to avoid these kinds of attacks.
\end{abstract}

\section{Introduction}
Satoshi Nakamoto proposed blockchain technology as a distributed ledger of connected records that are linked using cryptography measures \cite{b21}. The development of blockchain systems can be challenging even for experts because of the distributed execution environment and the persistence of records. Known consensus vulnerabilities increase the cost of errors. For example, the popular cryptocurrency exchange and wallet Coincheck got a security incident in January 2018 and more than 500 million USD were stolen \cite{b27}.

The model checking verification approach introduces methods to construct models that describe the possible system behavior in a mathematically precise way. The accurate modeling of systems often leads to the discovery of incompleteness in informal system specifications. One can write specifications in different formalisms: linear-temporal logic, computational tree logic, and extensions of them. Finite-state automata are the most expressed way to model systems.  The model checking process consists of three parts \cite{b78}:  
\begin{itemize}
  \item {\bf Modeling phase:} model the system; formalize the property to be checked.
  \item {\bf Running phase:} run the model checker engine to check the validity of the specifications.
  \item {\bf Analysis phase:} if the system satisfies the property, then check the next properties; otherwise, generate and analyze the counterexample, refine the model and repeat the entire procedure.
\end{itemize}

In this paper, we apply a statistical model checking approach to blockchain systems. We model three types of attacks that affect the major parties in blockchain systems. In the DNS attack, an adversary changes the DNS address for the connection to the network. The probability distribution function defines the address spoofing success.  Memory pool flooding and consensus delay attack models implement the scenario of double-spending of the same asset. Size of the memory pool and delay time one can take from the historical data of the Bitcoin network. Based on the experimental analysis, we consider approaches to reduce the probability of success of these attacks. The results of such analysis one can use in the planning of industrial product development.

The rest of the paper is organized as follows. In section \ref{sec:II} we describe tools and approaches that will be used in model checking. Section \ref{sec:III} explains attack scenarios and their models and provides restrictions on them. We carry out experiments in section \ref{sec:IV}. In section \ref{sec:V} we evaluate experimental results and suggest solutions for the prevention of successful attacks. In section \ref{sec:VI} we provide related studies on the modeling of smart contracts. Section  \ref{sec:VII} provides a discussion and the concluding remarks.

\section{Background} 
\label{sec:II}
In this section, we describe a tool and techniques that we use to construct models. We also illustrate the work of tools with a simple example. We use BIP (Behavior, Interaction, Priority) framework and a statistical model checker SBIP 2.0 \cite{b33} to model adversary's scenario on Blockchain. The stochastic real-time BIP formalism provides an ability to build models assembled from components presented as stochastic timed automata. Timed automaton is an extension of ordinary automaton by a finite set of clocks and clock constraints. One can think about stochastic timed automaton as a combination of timed automaton and Markov chain. Probability density functions can describe uncertainty in the model.  A designer of the system can provide a specification of properties as a Metric Temporal Logic (MTL) formula \cite{b36}. BIP framework uses a statistical model checking approach \cite{b33} that answers two types of questions:
\begin{itemize}
  \item {\bf Qualitative:} is the probability for the stochastic system $S$ to satisfy $\phi$ greater or equal to a certain threshold $\Theta$? The approach to answering the qualitative question is based on hypothesis testing.
  \item {\bf Quantitative:} what is the probability for $S$ to satisfy $\phi$? Given a precision $\delta$ and a risk parameter $\alpha$: $\mathcal{P}(|p'-p| < \delta) \geq 1 - \alpha$ the algorithm computes a value for $p'$.
\end{itemize}

In BIP, systems consist of three parts: atomic components, component interaction, and priority of interactions. Below we describe each of these parts.

\subsection{Atomic Components}
Timed automaton specifies the behavior of atomic components. One can define an atomic component as the following elements:

\begin{itemize}
  \item set of ports for synchronization with other components
  \item set of states of the component
  \item set of variables for local data
  \item set of transitions between states; the transition executes under certain boolean conditions.
\end{itemize}

\begin{figure}[htbp]
\centerline{\includegraphics[width=88mm,scale=0.5]{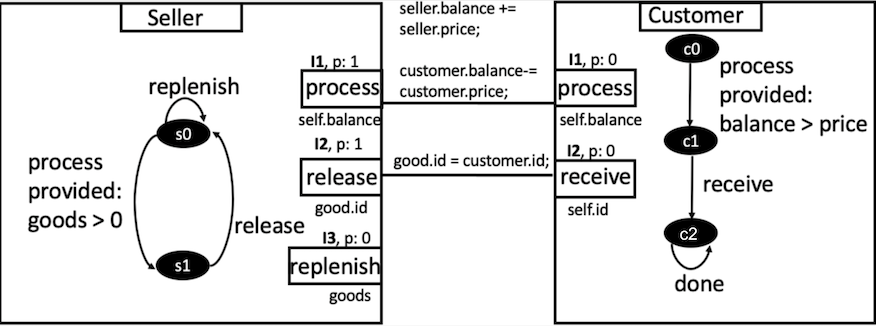}}
\caption{BIP model example}
\label{fig1.1}
\end{figure}

Fig.~\ref{fig1.1} illustrates a simplified purchasing model \cite{b52}, adapted for the BIP framework.  The atomic component \textit{Customer} has 3 states \textit{c0}, \textit{c1}, and \textit{c2}, variables \textit{balance}, \textit{transfer}, \textit{price} and \textit{id}. Also, the \textit{Customer} component has two ports \textit{process} and \textit{receive}, and three transitions: \textit{process}, \textit{receive} and \textit{done}.  Interaction \textit{process} executes only when the local variable \textit{balance} $>$ \textit{price}. The other atomic component explains the behavior of the seller with 2 states, \textit{s0}, \textit{s1}, internal variable balance and list of goods, and three transitions. The transition \textit{process} executes only when the number of goods is greater than zero.

\subsection{Connectors and Interactions}
Atomic components can communicate with each other according to the logic in connectors. One can describe a connector as a sequence of ports that connect atomic components. SBIP 2.0 supports two types of interactions, timed and stochastic. Timed interactions take place only with time constraints that represent a lower and upper bound over clock valuations, as in timed automata. Stochastic interactions take place with a specific stochastic constraint, for example, a probability function.

Interaction \textit{I1} at Fig.~\ref{fig1.1} connects atomic component \textit{Customer} through the port \textit{process}. This port implements the money transfer from the \textit{Customer}  to the \textit{Seller} atomic component. Variable \textit{balance} on interaction \textit{I1} of the customer is reduced on the amount \textit{price}, and the variable \textit{balance} of the seller is incremented on the same amount. In interaction \textit{I2} the asset assigns to the customer. Thus, the connector from example at Fig.~\ref{fig1.1} is a set of ports: $p_1|p_2$, where $p_1$ - port of atomic component \textit{Customer} and $p_2$ - port of external atomic component \textit{Seller}.

\subsection{Priorities}
Execution of priorities can proceed under certain conditions. If the condition holds, the priority of the considered connector is higher than another one. In the purchasing example from Fig.~\ref{fig1.1} a conflict can be between interactions \textit{I2} and \textit{I3} when the amount of good is non-zero. The interaction \textit{process} executes first, as it has a higher priority.

\subsection{Compound Components}
Compound components are used for assembling a new component from the defined atomic components. Compound components include instances of atomic components and specify connectors between them. Fig.~\ref{fig1.1} illustrates a compound component that consists of two atomic components and two interactions.

\section{Models Description} 
\label{sec:III}
In this section, we describe models and attack scenarios. We study attacks from the recent survey \cite{b48} that possesses the most meaningful properties for blockchain systems:
\begin{itemize}
  \item Affect peer-to-peer system or blockchain application.
  \item Led to the significant funds leak in the past. 
  \item Involve several parts of the blockchain systems.
  \item Include non-deterministic interactions.
\end{itemize}

We present models for the DNS spoofing attack, double spending through the mempool flooding attack, and consensus delay attack. Each model describes a particular attack scenario. We focused on one probabilistic parameter in each model. It is either a probability distribution or a historical data set. Particular transition in the model depends on this probabilistic parameter. The code of the SBIP models is available in the repository \cite{b56}.

\subsection{DNS Attack}\label{AA}
Blockchain applications use peer-to-peer network architecture for communication between the network nodes. When a new client joins a blockchain for the first time, she discovers active peers using Domain Name System (DNS) for IP address resolution. This is a DNS bootstrapping process.

The DNS mechanism is susceptible to cache poisoning, hijacking, tunneling, man-in-the-middle, and other types of attack \cite{b3}. A vulnerability in DNS led to the famous 2016 cyberattack \cite{b49}.  

We consider a DNS cache poisoning attack performed with a co-called birthday attack \cite{b3} on Berkeley Internet Name Domain (BIND) software. A BIND server can send multiple simultaneous recursive queries for the same IP address so an adversary can predict the next transaction id. After guessing the next transaction id, the adversary provides extra information in a DNS reply packet. Thus, by simultaneously generating a flow of queries to the server and an equal number of forged replies one can get a collision of transaction id and change the proper domain address to the fake one. The probability of collision is $P = 1 -(1- \frac{1}{t})^{\frac{n*(n-1)}{2}}$ \cite{b3}, where t is the total number of possible values in the master set, and n is the number of spoofed queries. The default number of possible values in the master set is $65535$. 

We took a birthday attack because the probability of collision is much higher than in conventional spoofing. The collision probability quickly increases up to $1.0$ along for less than $1000$ requests \cite{b3}.

\begin{figure}[htbp]
\centerline{\includegraphics[width=85mm,scale=0.7]{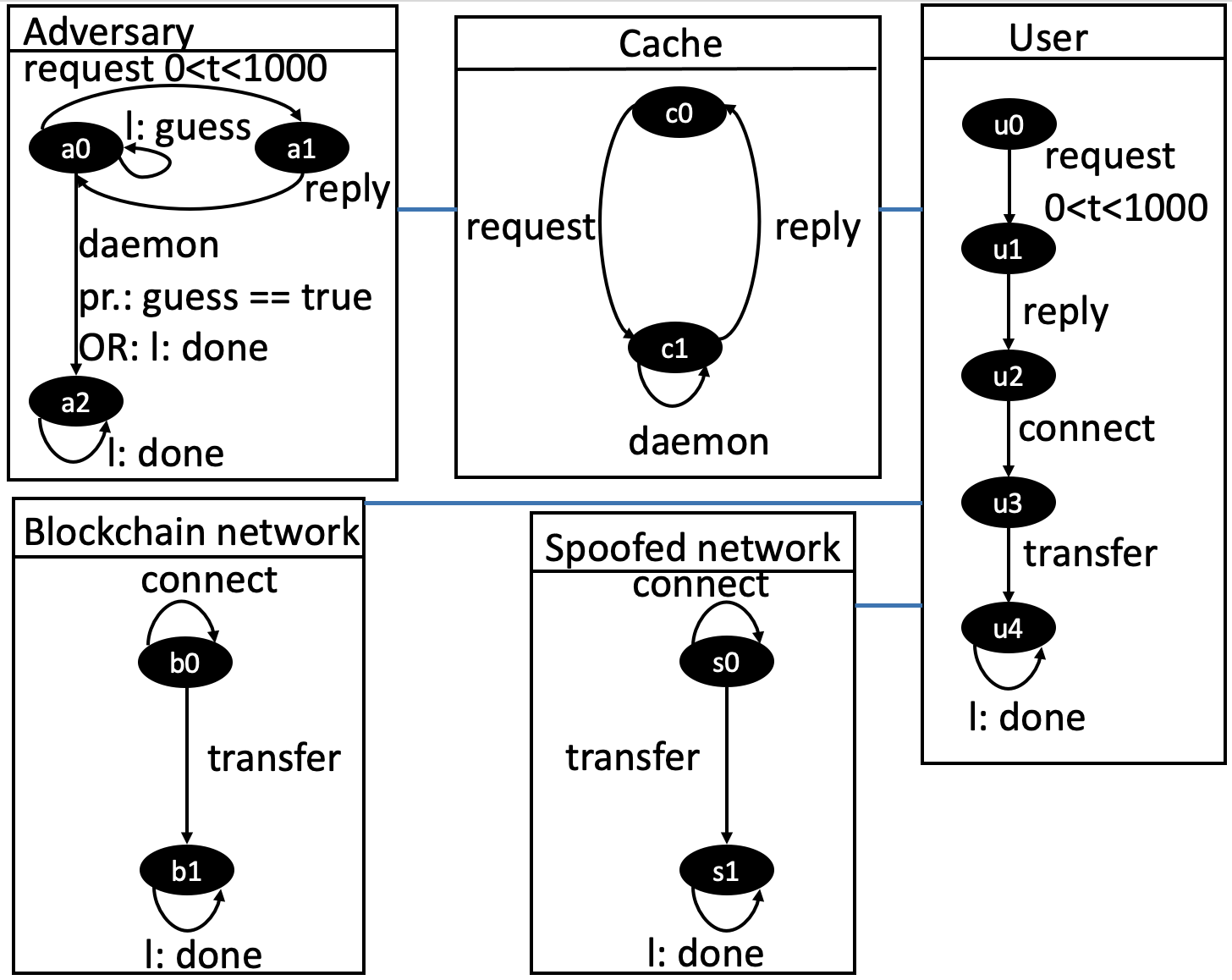}}
\caption{DNS spoofing model}
\label{fig2}
\end{figure}

Let us describe an SBIP model of the DNS cache poisoning birthday attack. Fig.~\ref{fig2} illustrates the model. Here and on the other models, transitions with the prefix \textit{l} present local interactions. The blue line indicates a connection between two atomic components. Connectors apply to interactions with the same name. For example, the adversary's atomic component can communicate with cache through external ports \textit{requests}, \textit{reply}, and \textit{daemon}.  The behavior of the atomic components is the following:
\begin{itemize}
	\item Adversary. Provides a set of requests and replies. After each request, he tries to guess a transaction ID. A random event here is the \textit{guess} transition, described by a collision probability function \cite{b3}. If the guess is successful, the adversary runs a daemon through the external interaction and goes to the final state $a2$.
	\item Cache. The cache server replies to DNS queries. In the case of collision, the interaction \textit{daemon} works. The address in the DNS cache changes, and the user gets a spoofed address. 
	\item User. Requests the DNS address from the cache server, connects to the network, and transfers funds. In case of a spoofed address, the transfer proceeds to the wrong recipient.
	\item Blockchain network. Awaits until the user connects to it and accomplishes transfer funds.
	\item Spoofed network. The same functionality as a blockchain network, but with the spoofed address.
\end{itemize}

The scenario of the attack is the following. The adversary makes requests to change the DNS address. At the same time, the user makes a query to the server. The request from the user and the adversary are temporary interactions that accomplish in the time period [0, 1000]. The upper bound of the period is a parameter, and one can change it. With the received address the user connects to the network and transfers the money.

One can bound the probability of the birthday attack by a negligible function $\mu$ by using cryptographic encryption techniques, for example, DNSSEC with NSEC5 records bounds \cite{b4}. NSEC5 is a resource record that can be used to detect certain attacks on secure DNS requests. We restricted the probability of violation by the function  $\mu = R*q_s*2^{-n}$ \cite{b4}, where $R$ - the set of domain names, that equals 65535,  $q_s$ - the number of adversary's queries, $n$ - the length of the output of the hash function. Restriction applies on the transition \textit{daemon} of the \textit{Cache} atomic component. 

\subsection{Double Spending and Mempool Flooding}\label{AB}
\begin{figure}[htbp]
\centerline{\includegraphics[width=87mm,scale=0.6]{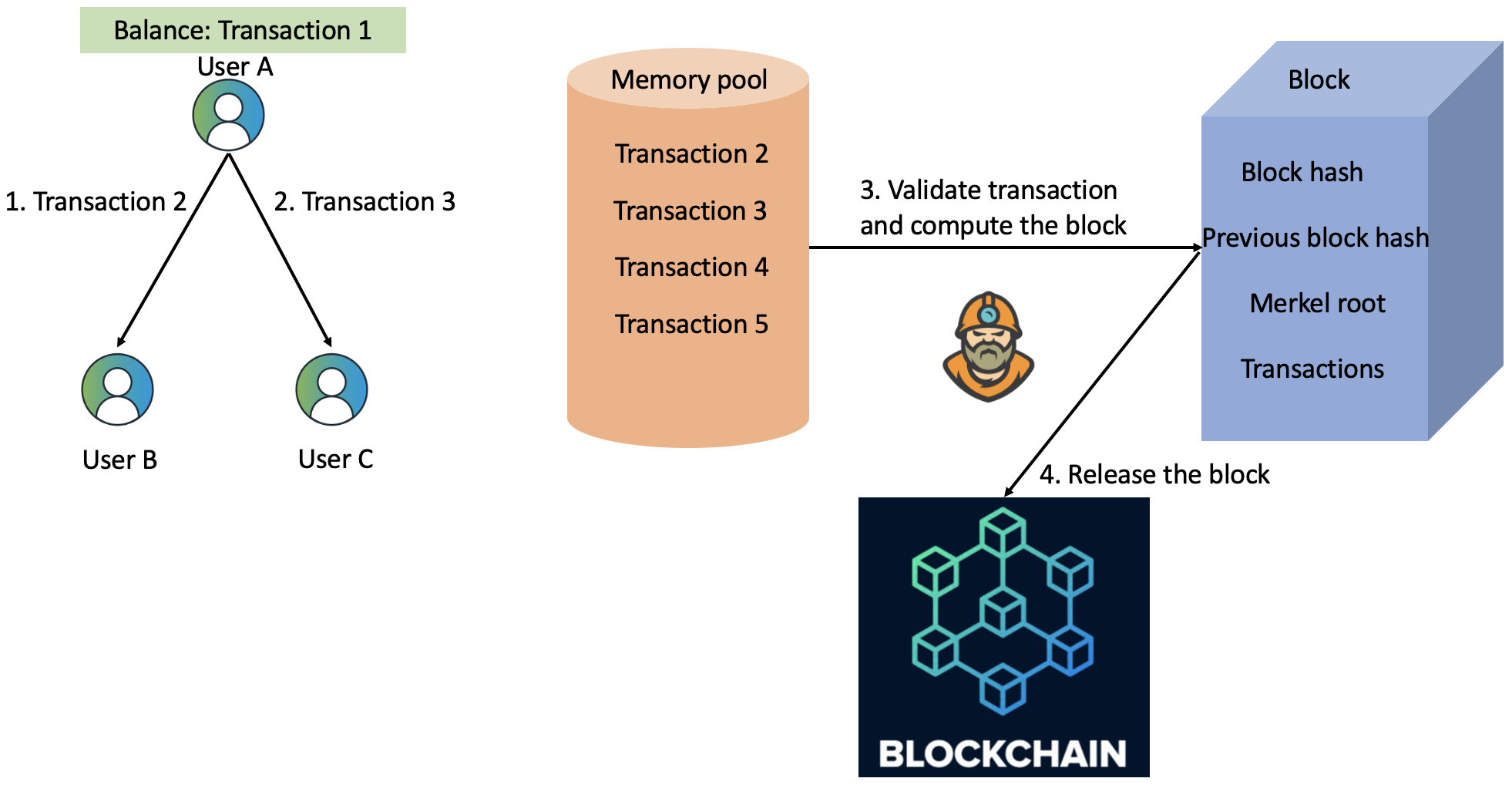}}
\caption{Double spending scenario}
\label{fig4}
\end{figure}

In a blockchain network, double-spending means applying the same transaction and its asset multiple times. Fig.~\ref{fig4} illustrates the process of double-spending.  User A signs the transaction with a private key and sends it to user B. During the validation process, the recipient looks up the unspent transactions of the sender, verifies the sender’s signature, and waits for the transaction to be mined into a valid block. Releasing the transaction to the network takes a certain time, which depends on the network throughput, the size of the memory pool of unconfirmed transactions (mempool), the consensus protocol, and the priority factor of the transaction.

In the case of fast transactions, the receiver can release the asset to the sender before the transaction gets committed into the blockchain network. In a while, user A can send the same transaction to user C. In this case, only those transaction is valid which gets into the blockchain first, but the asset might have been released twice. Double spending is one of the most widespread attacks on the blockchain platforms \cite{b54}. We performed a model checking for different types of double-spending attacks.

An adversary can cause a delay for successful double-spending with a mempool flooding attack. Mempool flooding is a kind of DDoS attack carried out at the memory pools of the cryptocurrencies \cite{b54}. Mempool has substantial properties such as timeout for transactions and default size limit. Relying on these configurations, users can estimate it to prioritize transactions. A potential adversary can estimate a delay in the mempool queue and increase this delay by a DDoS attack.

The survey \cite{b74} proposes two approaches to avoid double-spending. First, to set up the listening period before delivering the asset to the adversary. That gives time to double-check the spoofed transaction if the adversary already sends it to another user. But this does not consider the worst-case scenario when the adversary sends the second transaction just before accepting the first one to the blockchain.  The second approach implies insertion to the network additional "observers" - nodes that would relay to the user all transactions that it receives. But observers entail additional upload to the network. Instead, we propose to set the time interval in which the blockchain participant can send one transaction to the network.

\begin{figure}[htbp]
\centerline{\includegraphics[width=87mm,scale=0.6]{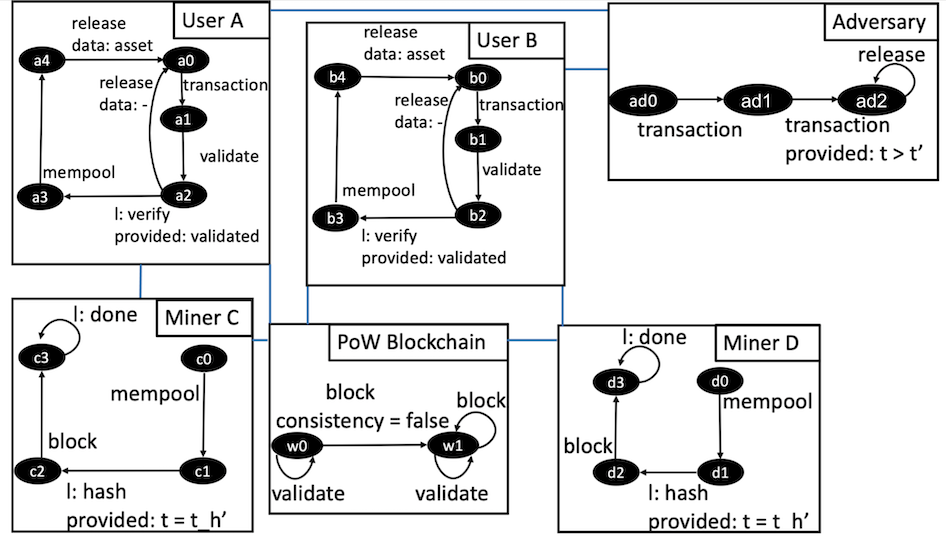}}
\caption{Mempool flooding model}
\label{mempool}
\end{figure}

Model on Fig.~\ref{mempool} represents a simplified version of the mempool flooding scenario. We neglect the probability that one can send two transactions to the same miner by assuming that the blockchain network consists of a significant amount of participants. Also, in our model, there is no explicit mining fee for transactions. The historical data set considers the notion of the different fees for transactions implicitly. Below we explain atomic components.

\begin{itemize}
\item Adversary. Sends a transaction from the same parent's block to the different recipients. After that, the adversary waits for the release of the transaction's assets.
\item Users A, B. Get the transaction from the sender, validate and locally verify it. Validation implies consistency with the blockchain's history and verification examines the sender's signature. After the verification phase, the user sends the transaction to the mempool and immediately releases the asset. The double-spending is successful if both assets were released to the sender.
\item Miners C, D. Take transactions from users and allocate them in the local mempool until the block will be mined. Mining time is a random variable with distribution from the historical dataset  \cite{b51}. 
\item Proof-of-Work (PoW) Blockchain. We presented all other nodes as a \textit{PoW Blockchain} atomic component. It takes the block from the miner, checks that the block does not contradict the history of committed transactions, and based on this either accepts or rejects the block.
\end{itemize}

The adversary sends a transaction to user A. After some time, the model defines a lower bound as a $t'$ parameter, the adversary generates a transaction with the same history as the first one and sends it to user B. When the user receives the transaction, she validates its consistency with the blockchain's history. If the transaction is consistent, the user verifies the sender's signature and sends it to the mempool. The user releases a product to the sender before transaction confirmation. 

\subsection{Double Spending and Consensus Delay}\label{AC}
Another way to enlarge the accepting time of the transaction is a consensus delay attack \cite{b14}.  While the memory pool increases the delay of a block because of the transactions queue, consensus delay appears because of the propagation time among the majority of users.  When a new block is ready, the majority of blockchain's participants should confirm it.  From the example in Fig.~\ref{fig4}, only one transaction gets acceptance which is the first to propagate among more the half of the whole nodes.

\begin{figure}[htbp]
\centerline{\includegraphics[width=87mm,scale=0.6]{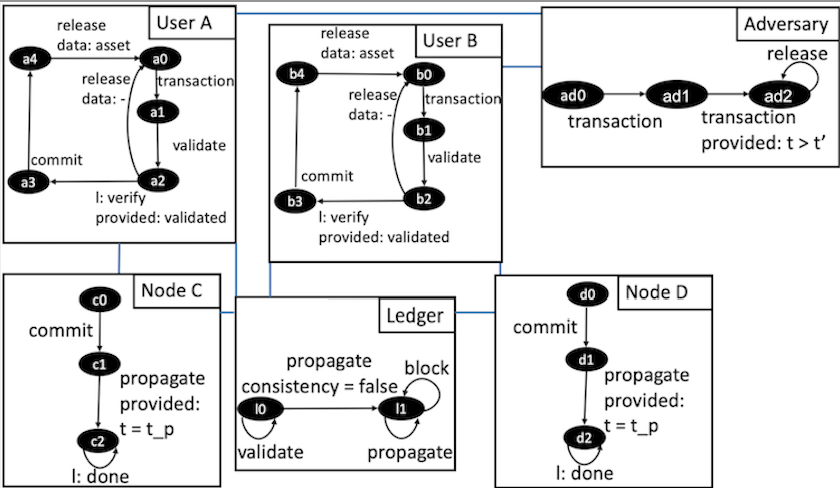}}
\caption{Consesus delay model}
\label{delay}
\end{figure}

Fig.~\ref{delay} illustrates the consensus delay model with the Proof-of-Work consensus protocol. The behavior of the adversary and users are the same as in the mempool flooding model. When a node gets transactions from the user, it builds them into the block and propagates this block among other nodes. Propagation takes some time and causes the delay. \textit{Ledger} atomic component represents other nodes of the network. 
 
In the consensus delay model, the adversary sends two transactions at a time interval $t'$. Users validate and verify transactions and send them to other nodes to propagate them. If the first transaction already got commitment by the network, the second benign user rejected the transaction on the \textit{validate} transition. We fixed the time interval $t'$, in which the adversary can send the spoofed transaction.

\section{Experimental Evaluation} 
\label{sec:IV} 
We implemented three models with SBIP 2.0 framework and estimated the probability of satisfying the specification. In each model, we introduced a probabilistic parameter. In a DNS attack, it is a probability of collision, in mempool flooding, it is the size of the memory pool. In the consensus delay model, the probabilistic parameter is a block propagation time. Each run can be either successful or not for an adversary, based on the probability distribution. The final ratio of successful attacks defines the probability of a successful adversary's scenario. 

In the current section, we provide probability success estimated with the SBIP tool. The goal of experiments is to get the rate of the adversary's success and analyze the result to reduce the success probability of the adversary. We take time as a parameter of each experiment, as it is the most important factor that affects the result.  Further, for each model, we propose a way to decrease the probability of adversary success.  SBIP tool runs the model a certain number of times, the experiment parameters define this number. Parameters for the tool are the following: $\delta = 0.1, \alpha = 0.1$. If we take them less than $0.1$, then the precision of the quantitative result of the experiment does not change significantly, but the evaluation time grows. In experiments we use statistical model checking algorithm \cite{b33}, based on probability estimation method \cite{b35}.  The code of models one can find in the repository \cite{b56}.

Experiments imply the following assumptions. Adversary makes a DNS attack on BIND software with the corresponding probability distribution of the collision \cite{b3}. In the double-spending model, we consider a fast-transactions network. We also emphasize the reliability of measurements. If statistical data is not objective, i.e. there is a gap or multiple null values, then measurements are not reliable. If the experiment violates these assumptions, then the result is not correct.

The experiments were run on a 2,3 GHz Dual-Core Intel Core i5 CPU. The time and memory limits are 90 minutes and 4 GB, respectively.

\subsection{DNS Attack}
In this subsection, we provide the estimation of the adversary's success probability for the DNS model. We calculated a discrete set of values of the probability collision function. According to the MTL specification, the balance of the spoofed network eventually becomes nonzero: $F[0, x] {spoofed.amount > 0}$, where $x$ is a parameter of the time in the range [100, 1000].

\begin{figure}[htbp]
\centerline{\includegraphics[width=79mm,scale=0.75]{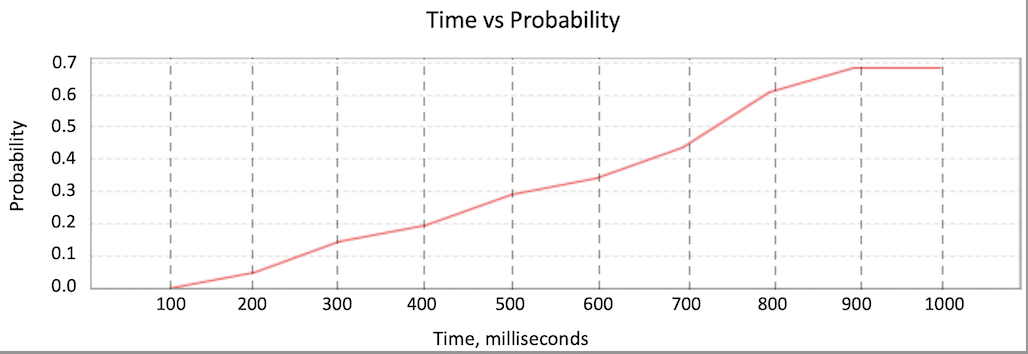}}
\caption{Dependence of the probability on the time for the DNS attack model}
\label{dns_probability}
\end{figure}

Fig.~\ref{dns_probability} depicts a probability versus time plot. Time denotes the number of milliseconds from the start of the scenario.  If we add the probabilistic interaction, that restricts the collision success, the probability of transfer to the wrong network decreases significantly. We checked the violation probability with the different hash functions with different lengths: SHA-1, SHA-224, SHA-256, SHA-384, SHA-512 \cite{b7}. The success of the \textit{daemon} interaction depends on the probability function $\mu = R*q_s*2^{-n}$ \cite{b4}, which we over-approximated by the uniform distribution.

\subsection{Double Spending and Mempool Flooding}
Based on the Bitcoin data from 01/01/2016 till 01/05/2021 we provide the mining time \cite{b51} as a stochastic variable in our SBIP model. The stochastic variable $t_h$ is assigned to the value from the dataset uniformly and randomly. 

The experiment aims to get a minimum time interval $t'$ in which the same user can send two transactions with the negligible probability of double-spending. The specification here represents the success of getting assets from both users: $F[0, 10^{12}] adversary.asset == 2$. We took such an upper bound for the specification to over-approximate the mining time from the historical data set. 

\begin{figure}[htbp]
\centerline{\includegraphics[width=77mm,scale=0.6]{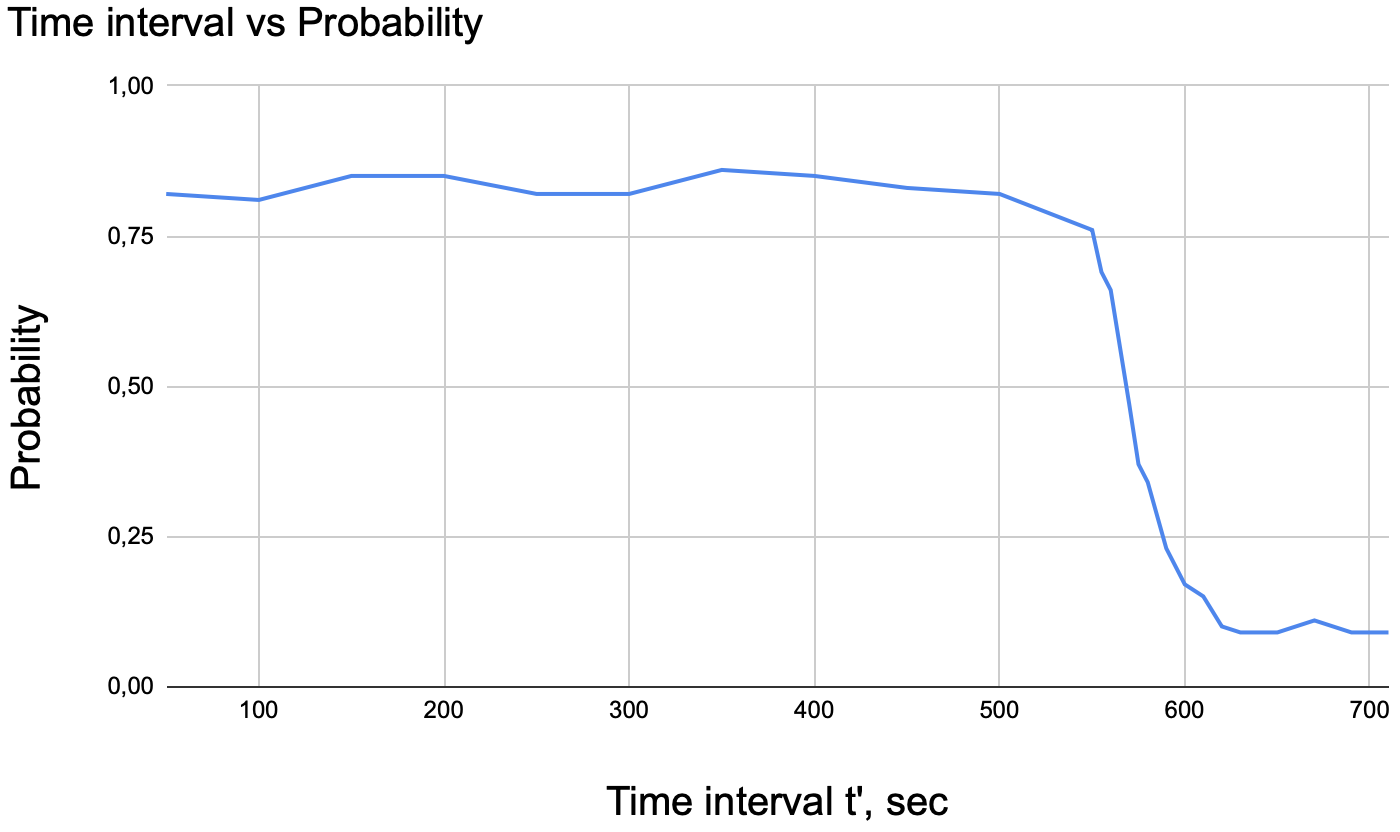}}
\caption{Dependence of the probability on the time interval $t'$ for the mempool flooding model}
\label{mempool_probability}
\end{figure}

Fig.~\ref{mempool_probability} illustrates the dependency of the double-spending probability on the time interval $t'$. We estimated the double-spending success probability with the time step in 10 seconds for $t'$ between 500 and 650. Between 0 and 500, we measured with the time step in 50 seconds, as the probability changes more slowly. For the time interval, $t'$ between 0 and 500 milliseconds, the probability of the double-spending is around 0.8. The following characteristics of the experiment explain this value. First, the value of the precision and risk parameters in the experiment is $0.1$. Second, the adversary can send a transaction later than in $t'$ seconds, as we provided the lower bound for the time interval.

\subsection{Double Spending and Consensus Delay}
Similar to the mempool double-spending scenario, we used the data \cite{b20} of the block propagation time to get a stochastic variable value. The goal of the experiment is to get the time interval $t'$ between the adversary's transactions with which the double-spending probability is negligible.  We took the historical data of propagation delay from 01/29/2016 till 01/05/2021 and assigned this data set to the random variable \textit{t'} of the atomic component \textit{Node}.

\begin{figure}[htbp]
\centerline{\includegraphics[width=77mm,scale=0.6]{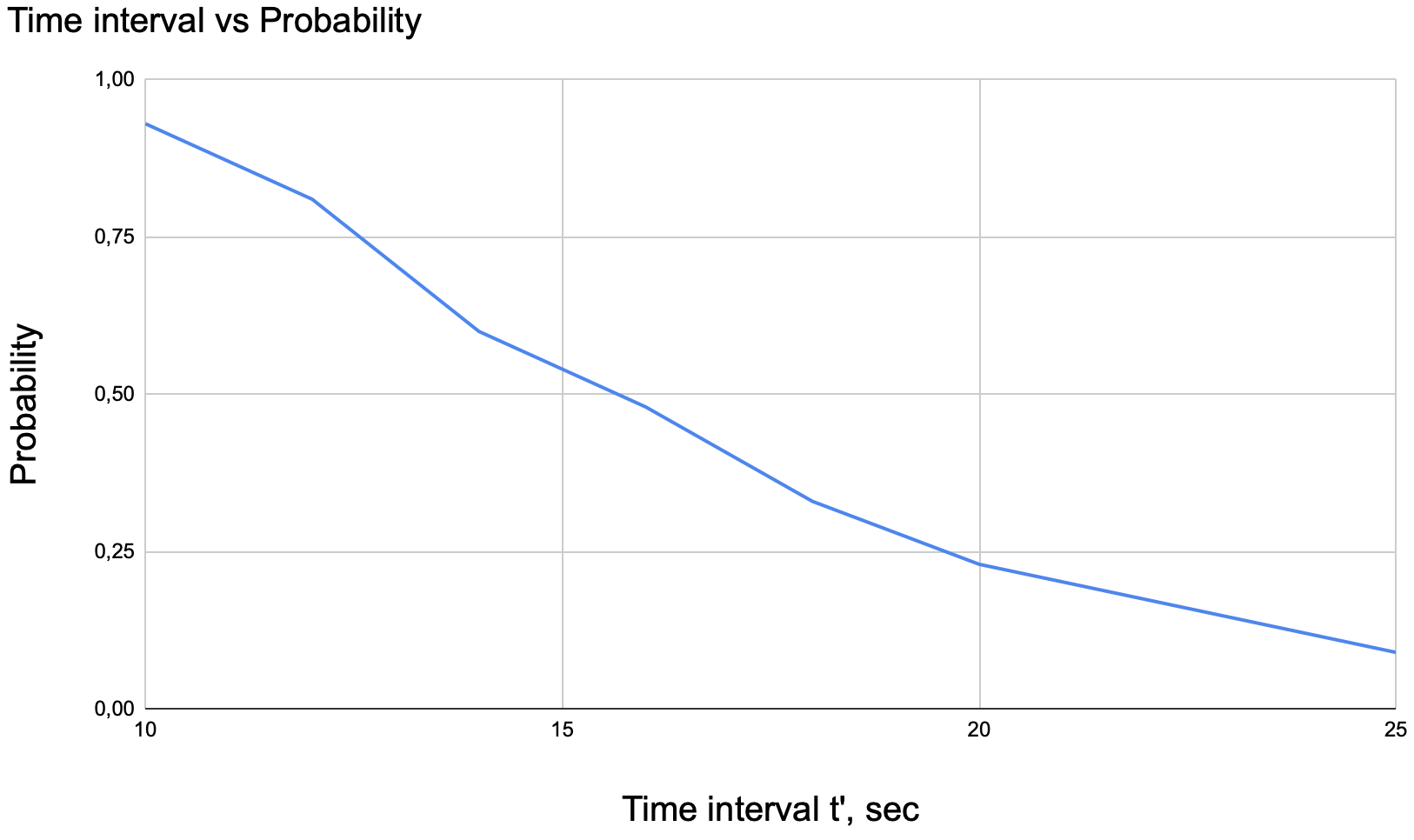}}
\caption{Dependence of the probability on the time interval $t'$ for the consensus delay model}
\label{delay_probability}
\end{figure}

The specification checks that the adversary got the double-spending of the same transaction: \\ $F[0, 10^{12}$] $adversary.asset == 2$. We accomplished measurements with the time step equals to 2 seconds. We over-approximated the propagation time in the specification to include all possible delays.  Fig.~\ref{delay_probability} illustrates the dependency of the time interval $t'$ from the double-spending probability.

\section{Analysis of Experimental Results}
 \label{sec:V}
 In this section, we estimate the results of experiments and evaluate proposed solutions to decrease the probability of a successful attack. A user can refine models with additional elements that specify her network. These elements can be a connection delay, network topology, bots for tracking DDoS attacks, and others. It is worth mentioning that a historical data set can include these elements implicitly. In this case, there is no need to change the model. Thus, one can use our models in their blockchain network.
 
\subsection{DNS Attack}
As we have seen from the experiment, the probability of the connection to the wrong network grows with the number of requests from the adversary. The probability function illustrated in Fig.~\ref{dns_probability} is similar to the collision rate function \cite{b3}. That means that the experimental evaluation corresponds to the theory. One can see from the experiment that for hash functions with an output length of more than 20 bits, the collision probability equals zero. Thus, using any
listed hash functions \cite{b7}  prevents the adversary’s scenario. 

\subsection{Double Spending and Mempool Flooding}
For the mempool flooding model, we analyzed the impact of the time interval between two transactions on the double-spending success. The increment of the time interval $t'$, in which the adversary can send a transaction, decreases the double-spending probability. After 570 seconds, the probability of successful double-spending becomes less than 0.5. After 610 seconds, the probability becomes around 0.1 and remains in this value asymptotically. That follows from the setting of experimental parameters $\delta$ and $\alpha$ to 0.1 value. Also, the average time in the mempool, which is around 600 seconds, corresponds to the measured time interval, with which the double-spending probability becomes negligible. The double-spending probability should decrease significantly around average mining time, as after this point the first transaction most probably releases from the mempool. According to the dependency illustrated in Fig. ~\ref{mempool_probability}, by setting the restriction for sending one transaction from the same user in 650 seconds one can reduce the probability of the double-spending success significantly. 

 Justification of the time restriction on transactions from the same user depends on the application area. That makes sense in the networks, for which confirmation and release asset of the first transaction is more crucial than the second transaction processing. 

\subsection{Double Spending and Consensus Delay}
Changing of probability versus time function in the consensus delay scenario, is smoother than in the mempool case, as there is a higher deviation in the data set. Fig.~\ref{delay} illustrates this dependency. Taking into consideration the result of the mempool flooding experiment, we conclude that by restricting the time interval $t'$ between two transactions from the same user in 650 seconds one can avoid both mempool flooding and consensus delay double-spending scenarios. To cover all factors that increase the double-spending success, one needs to make transaction fees and the mempool timeout as constants for all blockchain participants. The fixed transaction fee is possible to implement for the blockchain network \cite{b72}. One can fix the mempool timeout by assigning the same value for a timeout in memory pool configuration files for all blockchain participants and making these files immutable.

To sum up the experiment analysis, we specify security properties, that should hold for each system. In the case of a DNS attack, the system is safe if one uses any of the hash functions \cite{b7}. In the case of double-spending, one can put a time restriction for sending a transaction from one node. That bounds the successful attack probability by a negligible value and makes the system safe.

\section{Related Work} 
\label{sec:VI}
The problem of the statistical model is fundamental and well-studied \cite{b38}. One can use the statistical model checking approach to verify blockchain systems \cite{b45}. To the best of our knowledge, only two studies \cite{b46} \cite{b47} provide a statistical model checking evaluation of blockchain systems. But these studies do not present experiments on real attack scenarios and discussion of model restrictions. In our setting, real attacks with probability conditions and restrictions of the models have been considered.

\section{Conclusion} 
\label{sec:VII}
We modeled blockchain systems with historical data and probabilistic parameters. By running experiments on the historical data, one can predict the behavior of blockchain networks. We analyzed experimental results and proposed solutions to avoid the adversary's success and loss of money. Our models rely on real-world attack scenarios and cover all sides of the blockchain systems: miners, mining pools, exchanges, applications, users. Though we used open-source data from the Bitcoin network, one can adjust models for the private blockchain networks. Unlike in existing studies, our technique takes into account both temporal and stochastic interactions in blockchain systems.

As future work, we plan to expand the technique to automatically generate the SBIP models from the code of smart-contract. For some components of the blockchain system, such as memory pool or DNS cache it's difficult to build models automatically. But atomic components, that are responsible for the business logic in smart contracts, can be constructed in automatic mode. That extension reduces the manual work.

\bibliographystyle{eptcs}
\bibliography{references}
\end{document}